\documentclass[aps,prd,nopacs,nofootinbib,notitlepage,superscriptaddress,twocolumn]{revtex4-1}

\pdfoutput=1

\usepackage{amsfonts}
\usepackage{amsmath}
\usepackage{xcolor}
\usepackage[colorlinks=true,hyperfootnotes=true,citecolor=cyan]{hyperref}

\newcommand{\diff}{\textrm{d}}
\newcommand{\ddt}{\frac{\diff}{\diff t}}
\newcommand{\Q}{\mathbb{Q}}
\newcommand{\Qcg}{\mathring{Q}}

\begin{document}

\title{Pathological Character of Modifications to Coincident General Relativity: Cosmological Strong Coupling and Ghosts in $f(\Q)$ Theories}

\author{D{\'e}bora Aguiar Gomes}
\email{debora.aguiar.gomes@ut.ee}
\affiliation{Laboratory of Theoretical Physics, Institute of Physics, University of Tartu, W. Ostwaldi 1, 50411 Tartu, Estonia}

\author{Jose Beltr{\'a}n Jim{\'e}nez}
\email{jose.beltran@usal.es}
\affiliation{Departamento de Física Fundamental and IUFFyM, Universidad de Salamanca, E-37008 Salamanca, Spain}

\author{Alejandro Jim\'enez Cano}
\email{alejandro.jimenez.cano@ut.ee}
\affiliation{Laboratory of Theoretical Physics, Institute of Physics, University of Tartu, W. Ostwaldi 1, 50411 Tartu, Estonia}

\author{Tomi S. Koivisto}
\email{tomi.koivisto@ut.ee}
\address{Laboratory of Theoretical Physics, Institute of Physics, University of Tartu, W. Ostwaldi 1, 50411 Tartu, Estonia}
\address{National Institute of Chemical Physics and Biophysics, R\"avala pst. 10, 10143 Tallinn, Estonia}

%%%%%%%%%%%%%%%%%%%%%%%%%%%%%%%%%%%%%%%%%%%%%%%%%%%%%%%%%%%%%%
\begin{abstract}
The intrinsic presence of ghosts in the symmetric teleparallel framework is elucidated. We illustrate our general arguments in $f(\mathbb{Q})$ theories by studying perturbations in the three inequivalent spatially flat cosmologies. Two of these branches exhibit reduced linear spectra, signalling they are infinitely strongly coupled. For the remaining branch we unveil the presence of seven gravitational degrees of freedom and show that at least one of them is a ghost. Our results rule out $f(\mathbb{Q})$ cosmologies and clarify the number of propagating degrees of freedom in these theories.
\end{abstract}

\maketitle

%%%%%%%%%%%%%%%%%%%%%%%%%%%%%%%%%%%%%%%%%%%%%%%%%%%%%%%%%%%%%%
\textbf{Introduction.} The alternative geometrical interpretations of General Relativity (GR) \cite{BeltranJimenez:2019esp} in terms of torsion and non-metricity have received considerable attention in recent years. Not only do they allow for novel and fresh looks at different aspects of gravity, but they have also seeded new avenues to explore departures from GR. Experience has shown that modifying GR is not an easy task and it is common to find that modified gravity theories fail already at the purely theoretical level before even considering their potential phenomenological applications. While some of these problems appear quite evidently from the very construction of the theories, some other pathologies can be more subtle and, hence, easy to overlook. Furthermore, since they do not necessarily make their presence apparent in certain scenarios, the seriousness of the problems they bring along can be missed and not properly tackled. This occurs for the class of modified gravity theories formulated as extensions 
of the Symmetric Teleparallel Equivalent of GR (STEGR) \cite{Nester:1998mp} where the geometry only has non-metricity. The purpose of this Letter is to put forward the problematic nature of these theories. For the sake of concreteness, we shall focus on the so-called $f(\Q)$ gravity \cite{BeltranJimenez:2019tme} that has received a considerable amount of attention for cosmology and black hole physics lately (see e.g. \cite{Barros:2020bgg,Ayuso:2020dcu,Bajardi:2020fxh,Frusciante:2021sio,Atayde:2021pgb}), but their fundamental problems seem to be underappreciated. These theories are known to be infinitely strongly coupled around maximally symmetric backgrounds \cite{BeltranJimenez:2019tme,BeltranJimenez:2021auj}, a feature that contributes to the unappealing nature of these theories for phenomenology. However, the most serious hazard for these theories is the expected presence of ghosts. We will explain why ghosts are expected to arise in theories within the symmetric teleparallel framework and we will show it explicitly for the perturbations of cosmological solutions within $f(\Q)$ gravity. In doing so we will also unveil the presence of seven propagating degrees of freedom (d.o.f.) for these theories, thus clarifying their physical content.

%%%%%%%%%%%%%%%%%%%%%%%%%%%%%%%%%%%%%%%%%%%%%%%%%%%%%%%%%%%%%%
\textbf{Symmetric teleparallelism.} The symmetric teleparallel geometries are characterised by a flat and torsion-free connection so it can be written as
\begin{equation}
    \Gamma^\alpha{}_{\mu\beta}=\frac{\partial x^\alpha}{\partial\xi^\rho}\partial_\mu\partial_\beta\xi^\rho
\label{eq:defGamma}
\end{equation}
with $\xi^\rho$ arbitrary functions. It is possible to choose coordinates such that $\mathring{\xi}^\alpha=x^\alpha$ so the connection trivialises. This is the so-called coincident gauge. If we have a metric $g_{\mu\nu}$, the connection is determined by the non-metricity
\begin{equation}
    Q_{\alpha\mu\nu}=\nabla_\alpha g_{\mu\nu}.
\label{eq:defQ}
\end{equation}
In the coincident gauge, the non-metricity reduces to $\mathring{Q}_{\alpha\mu\nu}=\partial_\alpha g_{\mu\nu}$, while in an arbitrary gauge we need to use the connection \eqref{eq:defGamma}. Obviously, $\partial_\alpha g_{\mu\nu}$ is not a tensor under arbitrary diffeomorphisms and the non-metricity $Q_{\alpha\mu\nu}$ defined in \eqref{eq:defQ} can be interpreted as the covariantisation of $\partial_\alpha g_{\mu\nu}$ achieved by introducing suitable St\"uckelberg fields, namely $\xi^\alpha$. We refer to  \cite{BeltranJimenez:2022azb} for a more detailed discussion as well as other possible St\"uckelbergisation procedures.

The connection \eqref{eq:defGamma} possess a global symmetry given by
\begin{equation}
    \xi^\alpha\to A^\alpha{}_\beta\xi^\beta+b^\alpha
    \label{eq:defsym}
\end{equation}
with $A^\alpha{}_\beta$ and $b^\alpha$ a constant matrix and vector respectively. This symmetry represents the freedom to perform an arbitrary global general linear transformation of the coordinates together with a global translation. Thus, any theory constructed in the symmetric teleparallel geometries will have a set of conserved charges associated to this global symmetry. In particular, the shift symmetry guarantees that the equation for the St\"uckelberg fields will adopt the form of a conservation law 
\begin{equation}        
    \partial_\mu\left[\frac{\partial \mathfrak{L}}{\partial(\partial_\mu\xi^\alpha)}-\partial_\nu\frac{\partial^2 \mathfrak{L}}{\partial(\partial_\nu\partial_\mu\xi^\alpha)}\right]=0,   
\end{equation}
with $\mathfrak{L}=\sqrt{-g}\mathcal{L}$ the Lagrangian density. Although the charges derived from these conservation laws are physically useful, they do not play any role in the dynamical content of the theories. As it turns out, it is convenient to promote the associated global symmetries to gauge symmetries with the corresponding Bianchi identities in order to ensure the viability of the theory \cite{BeltranJimenez:2019odq}. The problem with theories formulated in the symmetric teleparallel arena can be better diagnosed by going to the coincident gauge where the non-metricity reads $\Qcg_{\alpha\mu\nu}=\partial_\alpha g_{\mu\nu}$. Thus, if we have an arbitrary action $\mathcal{S}[g_{\mu\nu,}\Qcg_{\alpha\mu\nu}]$, we will have derivative self-interactions for the graviton that break diffeomorphisms and, in general, we will have up to 10 propagating dof's. Since we know that a spin-2 field can propagate at most 5 healthy dof's if it is massive and 2 if it is massless, we conclude that these theories will be prone to propagating ghosts (see e.g. \cite{VanNieuwenhuizen:1973fi,Alvarez:2006uu,Hinterbichler:2011tt,deRham:2014zqa}). From the perspective of the St\"uckelberg fields, the presence of ghosts is associated to Ostrogradski instabilities \cite{Woodard:2015zca} because $\xi^\alpha$ enters with two derivatives in the connection \eqref{eq:defGamma} so that, for a generic action, the corresponding equations of motion will be fourth order. This is nothing but a version of the Boulware-Deser ghost \cite{Boulware:1972yco}. In this respect, the appearance of Ostrogradski ghosts in the symmetric teleparallel framework is different from other scenarios where they appear due to pathological non-minimal couplings of additional fields that do not arise as St\"uckelberg fields (see e.g. \cite{BeltranJimenez:2020sqf,BeltranJimenez:2020sqf,Delhom:2022vae}). The STEGR evades this problem because its action is given by the non-metricity scalar:
\begin{equation}
    \Q \equiv \frac{1}{4}Q_{\mu\nu\rho}Q^{\mu\nu\rho}-\frac{1}{2}Q_{\mu\nu\rho}Q^{\nu\mu\rho}-\frac{1}{4}Q_{\mu\nu}{}^\nu Q^{\mu\rho}{}_\rho+\frac{1}{2}Q_{\mu\nu}{}^\nu Q_\rho{}^{\rho\mu}\,,
\end{equation}
which has some remarkable properties. In the coincident gauge, it still possesses a gauge (diffeomorphisms) symmetry ensuring that only two components of the metric are propagating dof's. Let us notice that these diffeomorphisms are different from the original ones present in all the symmetric teleparallel theories and which are exhausted in the coincident gauge. Another way of understanding this result is by noticing that the St{\"u}ckelberg fields only enter as a total derivative [see e.g. \eqref{eq:bgQ} below] so their equations of motion are trivial rather than fourth order. This is another way of saying that the St\"uckelberg fields are not necessary to {\it restore} covariance because the theory already is covariant. 

Since the non-metricity scalar gives rise to an exceptional theory, one could think that their non-linear extensions $f(\Q)$ could also have remarkable properties. This happens for the curvature formulation of GR whose action is given in terms of the Ricci scalar $R$ of the Levi-Civita connection. The non-linear extensions $f(R)$, that could be expected to suffer from Ostrogradski instabilities, are actually free of them \cite{Woodard:2006nt} and they simply propagate one additional scalar that can be ascribed to a conformal mode of the metric. The case of $f(\Q)$ is however different and one hint that things are worse is that the $f(\Q)$ theories break the diffeomorphisms in the coincident gauge or, in other words, they bring the St\"uckelberg fields back to life through derivative interactions, which means that they will have higher order equations of motion. Some problematic features of these theories have already been pointed out in the literature (including the presence of a ghost \cite{BeltranJimenez:2021auj}) and here we will extend those findings. Before delving into that, we will discuss the cosmological configurations that can be constructed in the symmetric teleparallel geometries.

%%%%%%%%%%%%%%%%%%%%%%%%%%%%%%%%%%%%%%%%%%%%%%%%%%%%%%%%%%%%%%
\textbf{Cosmological configurations}. We will adapt the results of \cite{Gomes:2023hyk} to construct the St\"uckelberg fields configurations for spatially flat cosmologies. We will use the general teleparallel reference frames realising the ISO(3) symmetry of the spatially flat cosmologies that are named the trivial branch and the non-trivial branches I and II in \cite{Gomes:2023hyk}.\footnote{
    The names trivial and non-trivial allude to the realisation of homogeneity in the different branches. While in the trivial branch the homogeneity corresponds to standard spatial translations, in the non-trivial branches the homogeneity is realised as a combination of the spatial translations combined with internal GL$(4,\mathbb{R})$ transformations for the frames \cite{Gomes:2023hyk}.} 
Since the symmetric teleparallel corresponds to the integrable frames, we need to impose such a condition. We will do it for the three spatially flat branches and we find the following configurations for the St\"uckelberg fields:
\begin{itemize}
    \item[(a)] Trivial branch: 
    \begin{equation}
    \xi^0=\xi(t),\quad\xi^i=\sigma_0x^i,
    \end{equation}
    with $\sigma_0$ a constant. The isotropy is realised by compensating the spatial rotations $x^i\to R^i{}_jx^j$ with internal transformations \eqref{eq:defsym} of the form $A^i{}_j=(R^{-1})^i{}_j$, while spatial translations $\vec{x}\to \vec{x}+\vec{x}_0$ are compensated with a shift $\vec{b}=-\sigma_0\vec{x}_0$.
    \item[(b)] Non-trivial branch I:
    \begin{equation}
        \xi^0=\xi(t)-\frac12\sigma_0\lambda|\vec{x}|^2,\quad\xi^i=\sigma_0x^i.
    \end{equation}
    Rotations are realised as in the trivial branch, but translations need a little more work to guarantee homogeneity of $\xi^0$. The internal transformation that restores homogeneity of that component is given by $A^0{}_i=\sigma_0 \delta_{ij}x^j_{0}$ together with a shift $b^0=\frac12 \sigma\lambda|\vec{x}_0|^2$.
    \item[(c)] Non-trivial branch II:
    \begin{equation}
    \xi^0=\xi(t),\quad\xi^i=\Big[\lambda\xi(t)+\sigma_0\Big] x^i.
    \end{equation}
    The homogeneity of $\vec{\xi}$ now requires $A^i{}_0=-\lambda x_0^i$ and the shift $\vec{b}=-\sigma_0\vec{x}_0$, while rotations are realised as in the other branches.
\end{itemize}
The three branches can be related by introducing the following parameterisation:
\begin{equation}
    \xi^0=\xi(t)-\frac{\alpha_{\rm I}}2\sigma_0\lambda|\vec{x}|^2,\quad \xi^i=\Big(\alpha_{\rm II}\lambda\xi(t)+\sigma_0\Big) x^i\,,\label{eq:3branches}
\end{equation}
so the trivial branch is obtained for $\alpha_{\rm I}=\alpha_{\rm II}=0$ and the non-trivial branches I and II correspond to $(\alpha_{\rm I}=1,\alpha_{\rm II}=0)$ and $(\alpha_{\rm I}=0,\alpha_{\rm II}=1)$, respectively. The trivial branch is recovered from the non-trivial branches in the limit $\lambda\to0$ and all the branches are characterised by a single function of time. In terms of the symmetries, the non-trivial realisation of homogeneity trivialises as we take $\lambda\to0$.

The above cosmological configurations will then be compatible with the symmetries of the FLRW metric described by the line element
\begin{equation}
    \diff s^2=-n(t)^2\diff t^2+a(t)^2 \delta_{ij}\diff x^i \diff x^j\,,
\end{equation}
with $n(t)$ and $a(t)$ the lapse and the scale factor respectively. The non-metricity scalar then reads
\begin{equation}
    \Q=\frac{6\dot{a}^2}{a^2n^2}+ \frac{3\lambda}{a^3 n}\ddt  \left[\alpha_{\rm I} \frac{ \sigma_0 an}{\dot{\xi}}-\alpha_{\rm II} \frac{a^3}{n}\frac{\dot{\xi}}{ \lambda \xi + \sigma_0}\right]\,, \label{eq:bgQ}
\end{equation}
so the St\"uckelberg fields contribute a total derivative to $\sqrt{-g}\,\Q$. Let us notice that the non-metricity scalar has a time-reparameterisation symmetry $t\to \zeta(t)$, $n\to  n/\dot{\zeta}$ that does not involve the St\"uckelbergs. If the action is linear in $\Q$, the $\xi$'s do not contribute, as expected, but for non-linear functions they will enter with second order derivatives so the connection equation will be fourth order as anticipated above. This is an explicit indication of the general argument given above that these theories will be prone to Ostrogradski instabilities or, in other words, it can be interpreted as a signal that a version of the Boulware-Deser ghost is present. Since they enter as a total derivative, one might hope for the appearance of some special features for the $f(\Q)$ theories that would tame the ghost. This does not seem to be the case since the $f(\Q)$ theory can be brought into a form close to an Einstein frame given by \cite{BeltranJimenez:2021auj}
\begin{align}
    \mathcal{S}
    &=\frac12\int\diff^4x\sqrt{-q}\Big[R(q)+6(\partial\varphi)^2+\mathcal{U}(\varphi)\nonumber\\
    &\qquad\qquad-2\big(q^{\alpha\beta}q^{\mu\nu}-q^{\alpha\mu}q^{\beta\nu}\big)\partial_\alpha\varphi\nabla_\beta q_{\mu\nu}\Big],
    \label{eq:Einsteinframe}
\end{align}
where $q_{\mu\nu}$ is a metric conformally related to $g_{\mu\nu}$, $\varphi$ is the conformal mode and $\mathcal{U}(\varphi)$ its potential, which is determined by the form of the function $f(\Q)$. We refer to \cite{BeltranJimenez:2021auj} for more details and a more extensive discussion. Here we only want to highlight that the kinetic term of the conformal mode enters with the wrong sign, thus suggesting that the theory propagates a ghost. The reasoning is that the kinetic term of the conformal mode will contribute a negative component to the diagonal of the kinetic matrix of the system. Since a kinetic matrix with a negative value in the diagonal cannot be positive definite, we conclude that the theory is prone to propagating at least one ghost. Furthermore, the second line in \eqref{eq:Einsteinframe} contains either non-diffeomorphism-invariant derivative interactions between $q_{\mu\nu}$ and the conformal mode in the coincident gauge or, equivalently, second order derivatives of the St\"uckelberg fields coupled to derivatives of the conformal mode. In both cases, this would signal another potential source of ghosts. We will confirm the expected presence of ghosts with an explicit calculation.

%%%%%%%%%%%%%%%%%%%%%%%%%%%%%%%%%%%%%%%%%%%%%%%%%%%%%%%%%%%%%%
\textbf{Cosmological non-viability of $f(\Q)$ theories.} For the sake of generality, we will include a matter sector with a canonical scalar field $\chi$ so our action is given by
\begin{equation}
    \mathcal{S}=\int\diff^4x\sqrt{-g}\left[-\frac12 f(\Q)-\frac12(\partial\chi)^2-V(\chi)\right],
\end{equation}
with $V(\chi)$ some potential for the scalar field. The equations of motion can be obtained from the minisuperspace approach by introducing our cosmological Ansatz and taking variations with respect to $n(t)$, $a(t)$, $\xi(t)$ and the homogeneous scalar field $\bar{\chi}(t)$. We have explicitly checked that the obtained equations coincide with the covariant equations. We will not give their general form, but it is useful to realise that the connection equations for the non-trivial branches read
\begin{align}
&\text{branch I:} \qquad   \ddt \left(\frac{\lambda\sigma_0 a n \dot{f}_{\Q}}{\dot{\xi}^2}\right)=0, \label{eq:BIconserv}\\
&\text{branch II:}   \qquad  \frac{\lambda}{\lambda\xi+\sigma_0}\ddt \left(\frac{a^3\dot{f}_{\Q}}{n} \right)=0,\label{eq:BIIconserv}
\end{align}
while the equation for the trivial branch trivialises. For the branch I we obtain a conservation equation due to the obvious shift symmetry $\xi\to\xi+c$ with $c$ a constant that is inherited from the general shift symmetry $\xi^0\to\xi^0+b^0$ of the full theory. Although this symmetry is not present in the branch II, it still adopts the form of a conservation law. Both equations trivialise for $f_{\Q\Q}=0$, as one would expect since this corresponds to GR. Furthermore, the trivialisation of the connection equation in this case leads to the usual Bianchi identities for the background that, in turn, lead to the standard conservation equation for the matter sector.
In the general case (for any of the three branches), we still have the Bianchi identities
\begin{equation}
    n \ddt \frac{\delta \bar{\mathcal{S}}}{\delta n}-\dot{a}\frac{\delta\bar{\mathcal{S}}}{\delta a}-\dot{\chi}\frac{\delta\bar{\mathcal{S}}}{\delta \chi}-\dot{\xi}\frac{\delta\bar{\mathcal{S}}}{\delta \xi}\equiv0,
\end{equation}
where $\bar{\mathcal{S}}$ represents the action after evaluating the cosmological configuration (minisuperspace approach).

The conservation equations \eqref{eq:BIconserv} and \eqref{eq:BIIconserv}  can be integrated once to give
\begin{equation}
    -\frac{2\lambda\sigma_0 a n \dot{f}_{\Q}}{\dot{\xi}^2}=J_{\text{I}} \qquad\text{and}
\qquad    \frac{\lambda a^3 \dot{f}_{\Q}}{n}=J_{\text{II}},\label{eq:J}
\end{equation}
with $J_{\text{I}, \text{II}}$ the conserved charges that characterise the solutions (the numerical factors have been introduced for convenience). Let us notice that, for generic $f$, the solutions with trivial charge $J_{\text{I,II}}=0$ are only compatible with either the trivial branch $\lambda=0$ or, else, they require a constant non-metricity scalar $\Q$. 

We will now turn our attention to the more interesting inhomogeneous perturbations. For a clear comparison of the three branches, we will work around the generic configuration \eqref{eq:3branches} in terms of $\alpha_{\rm I}$ and $\alpha_{\rm II}$. We will use a gauge with unperturbed St\"uckelberg fields so $\delta\Gamma^\alpha{}_{\mu\nu}=0$. Since this choice exhausts the diffeomorphisms freedom, we need to take into account all the metric perturbations and we will parametrise them in conformal time $\eta$ (for which $n=a$) as
\begin{align}
    \delta g_{00} &= - 2 a^2 \phi  \,,\\
    \delta g_{0i} &= a^2 (\partial_i B + B_i)\,,\nonumber\\
    \delta g_{ij} &=  a^2\left[ -2 \psi \delta_{ij}+ \partial_i\partial_j E + \frac12(\partial_{i}E_{j}+\partial_{j}E_{i})+ h_{ij}\right] \,\nonumber,
\end{align}
with $\delta^{ij}h_{ij}=0$ and $\partial_iB^i=\partial_iE^i=0=\partial_i h^{ij}$. The derivatives with respect to the conformal time will be represented with a prime. The matter sector will be perturbed as $\chi=\bar{\chi}(\eta)+\pi(\eta,\vec{x})$. Let us then proceed to computing the quadratic action for the different sectors in increasing order of difficulty. For the subsequent computations we will work in Fourier space, make extensive use of the background equations of motion and perform integrations by parts. We will use the same symbols for the Fourier transform of the perturbations.

The tensor sector is governed by
\begin{equation}
    \mathcal{S}^{(2)}_{\rm ten}=\frac12\sum_{\alpha=+,\times}\int\diff \eta\diff^3k\ a^2f_{\Q} \left[\vert h'_\alpha\vert^2-\left(k^2+ \frac{\alpha_{\rm I}J_{\rm I}\xi'}{a^2f_{\Q}} \right)\vert h_\alpha\vert^2\right],
\end{equation}
and its healthiness requires $f_{\Q}>0$ to avoid ghosts in all the branches. The tensor modes also acquire a mass due to the non-trivial background charge $J_{\rm I}$ which only contributes in the branch I. This mass is positive provided $J_{\rm I}\xi'>0$.

The vector sector contains $\vec{B}$ and $\vec{E}$. However, $\vec{B}$ is non-dynamical and can be integrated out, leaving the following action
\begin{equation}
    \mathcal{S}^{(2)}_{\rm vec}=\frac12\int\diff \eta\diff^3k\ \alpha_{\rm I}J_{\rm I}\xi'\left[ \frac{\vert\vec{E}'\vert^2}{1+\dfrac{J_{\rm I}\xi'}{a^2 f_{\Q}k^2}} - k^2 |\vec{E}|^2\right]\,. 
\end{equation}
In the UV regime with $k^2\gg\frac{J_{\rm I}\xi'}{a^2f_{\Q}}$, we have
\begin{equation}
    \mathcal{S}^{(2)}_{\rm vec,UV}=\frac12\int\diff \eta\diff^3k\ \alpha_{\rm I}J_{\rm I}\xi'\left[\vert\vec{E}'\vert^2-k^2\vert\vec{E}\vert^2\right],
\end{equation}
so the absence of ghosts in the vector sector for the branch I requires $J_{\rm I}\xi'>0$. For the trivial branch and the branch II, this sector trivialises which means that $\vec{E}$ becomes strongly coupled in those branches and this can be associated to the appearance of an accidental invariance under transverse diffeomorphisms. 

Finally, the scalar sector contains the non-dynamical perturbation $B$ that we can integrate out to obtain:
\begin{equation}
    \mathcal{S}{^{(2)}_{\rm scal}}=\frac12\int\diff \eta\diff^3k\ a^2 \left[ \Phi'\hat{\mathcal{K}}_{\rm s}\Phi'{}^\dagger+\Phi'\hat{\mathcal{N}}_{\rm s}\Phi^\dagger-\Phi\hat{\mathcal{V}}_{\rm s}\Phi{}^\dagger+\text{c.c.}\right]
\end{equation}
with $\Phi=(\phi,\psi,E,\pi)$ and c.c. stands for complex conjugate. Instead of giving the full expression of the matrices, we will simply quote the kinetic matrix determinant
\begin{equation}
    \det \hat{\mathcal{K}}_{\rm s} =\alpha_{{\rm I}}\frac{9\lambda^2\sigma_0^2f_{\Q}^2f_{\Q\Q}}{2a^2\xi'^2 \left[1-k^2 \dfrac{(\xi''-2\mathcal{H}\xi')^2}{J_{\rm I}\xi'^3}f_{\Q\Q}\right]}k^4,  \label{eq:Det}
\end{equation}
with $\mathcal{H}\equiv a'/a$. This determinant vanishes for both the trivial branch and the branch II, showing that they have a reduced linear spectrum and, thus, they contain infinitely strongly coupled modes. On the other hand, all the remaining scalar modes propagate for the branch I (with $J_{\text{I}}\neq0$). If we now take the UV, the determinant reduces to
\begin{equation}
    (\det{\mathcal{K}_{\rm s}})_{\rm UV}\simeq-\frac{9J_{\rm I}\xi'\lambda^2\sigma_0^2f_{\Q}^2}{2a^2(\xi''-2\mathcal{H}\xi')^2}k^2.
\end{equation}
Since we need $f_{\Q}>0$ for the gravitons to be non-ghostly and the stability of the vector sector requires $J_{\rm I}\xi'>0$, the scalar sector cannot be ghost-free, since we have $(\det{\mathcal{K}_s})_{\rm UV}<0$. This proves the advertised presence of at least one ghost in the cosmological spectrum of the $f(\Q)$ theories in this branch and concludes our demonstration of the non-viability of spatially flat $f(\Q)$ cosmologies. We have performed an analogous analysis in the Einstein frame \eqref{eq:Einsteinframe} and confirmed the presence of a ghost. In fact, this analysis allow us to show the presence of a ghost for the scalar-non-metricity theories $f(\Q,\chi)$, the only difference being that the potential in \eqref{eq:Einsteinframe} is now $\mathcal{U}=\mathcal{U}(\varphi,\chi)$, and we will give the details elsewhere.

%%%%%%%%%%%%%%%%%%%%%%%%%%%%%%%%%%%%%%%%%%%%%%%%%%%%%%%%%%%%%%
\textbf{Discussion.} In this Letter we have discussed why theories formulated in a symmetric teleparallel geometry are generically prone to be plagued by ghosts. This is a feature that makes these theories unappealing for phenomenological applications such as cosmology or black hole physics. In order to illustrate these problems, we have focused on the cosmology of spatially flat FLRW universes within the class of $f(\Q)$ theories, since they might have been expected to be exceptional very much like $f(R)$ theories among generic metric theories containing higher order Riemann terms. As a byproduct of our analysis, we have unveiled the presence of seven dynamical dof's associated with the gravitational sector in one of the branches, thus clarifying the issue of the number of propagating dof's in $f(\Q)$, which has remained elusive in previous Hamiltonian analyses \cite{Hu:2022anq,DAmbrosio:2023asf,Tomonari:2023wcs}. Let us note that seven in fact exhausts all the possible dof's in $f(\Q)$ since the shift is never dynamical in these theories (as we have corroborated in our analysis of the cosmological perturbations). An upper bound of seven dof's has also been suggested in \cite{DAmbrosio:2023asf}. Thus, the propagating dof's in the coincident gauge correspond to the six components of the spatial metric plus the lapse. This is once more a signal for the presence of the Boulware-Deser ghost. In the Einstein frame \eqref{eq:Einsteinframe} where both the lapse and the shift are non-dynamical \cite{BeltranJimenez:2021auj}, these seven dof's could be associated with the two tensorial modes of the graviton, the conformal mode and the four St\"uckelberg fields $\xi^\alpha$. Our results will be useful to perform the Hamiltonian analysis for the perturbations in the branch I to elucidate the Hamiltonian structure of the theories without performing the full Hamiltonian construction.

The presence of 7 dof's in the non-trivial branch I allows us to establish the existence of strong coupling problems in the other two spatially flat branches that have reduced linear spectra. In the branch containing the 7 dof's, we have explicitly demonstrated the impossibility of having a ghost-free linear spectrum, in agreement with our general argument. Our results thus show the non-viability of all the spatially flat $f(\Q)$ cosmologies.

Our main conclusion is that the exposed shortcomings of the symmetric teleparallel framework need to be tackled before theories beyond Coincident GR formulated in this framework can be claimed to lead to physically sensible applications in e.g. cosmology and/or black hole physics. Additionally, having unveiled the number of dof's in $f(\Q)$ our results will permit to easily diagnose strong coupling problems in other backgrounds.

To end this Letter, let us comment that the problems with the symmetric teleparallel framework will also arise in the extensions of the general teleparallel equivalent of GR \cite{BeltranJimenez:2019odq} where the torsion-free condition is dropped. As already explained in \cite{BeltranJimenez:2019odq}, the spectrum of those theories contain additional spin-2 fields as well as non-gauge invariant derivative self-interactions for the spin-2 modes that will propagate ghosts. To give a more positive ending, we shall conclude by stating that, despite not being suitable for physical applications, this framework does possess a series of compelling theoretical features that can be interesting to explore in more detail. Furthermore, the symmetric teleparallel landscape can also be deformed so it can accommodate viable physical models. A potentially interesting route would be, for instance, providing a geometrical foundation for the ghost-free derivative interactions constructed in \cite{Hinterbichler:2013eza}.

%%%%%%%%%%%%%%%%%%%%%%%%%%%%%%%%%%%%%%%%%%%%%%%%%%%%%%%%%%%%%%
%\vspace{0.2cm}
{\bf Acknowledgments:}
We would like to thank Sebastian Bahamonde for discussions and Francisco Jos\'e Maldonado Torralba for useful comments. This work was supported by the Estonian Research Council grant PRG356 “Gauge gravity: unification, extensions and phenomenology”, and by the  CoE program TK202 “Foundations of the Universe”. J.B.J. was supported by the Project PID2021-122938NB-I00 funded by the Spanish “Ministerio de Ciencia e Innovaci\'on" and FEDER “A way of making Europe”. A.J.C. is supported by the European Regional Development Fund through  the Mobilitas Pluss post-doctoral grant MOBJD1035.

\bibliography{NonTrivialCosmology}

\end{document}